# Active formation of Friedrich-Wintgen bound states in the continuum in dielectric dimerized grating borophene heterostructure

Xiao-Fei Yan,[1,2,3,†] Xin-Yang Wang,[1,2,†] Qi Lin,[1,4] Ling-Ling Wang[4] and Gui-Dong Liu,[1,4,*]

[1]School of Physics and Optoelectronics, Xiangtan University, Xiangtan 411105, China

[2]Hunan Engineering Laboratory for Microelectronics, Optoelectronics and System on a Chip, Xiangtan University, Xiangtan 411105, China

[3]School of Electronic and Information Engineering, Beihang University, Beijing 100191, China

[4]School of Physics and Electronics, Hunan University, Changsha 410082, China

[†]These authors contributed equally to this work.
*Corresponding author: gdliu@xtu.edu.cn

**Abstract:** The Friedrich-Wintgen bound state in the continuum (FW BIC) provides a unique approach for achieving high quality factor ($Q$-factor) resonance, which has attracted wide attention and promoted the development of various applications. However, the FW BIC is usually considered as *accident* BIC resulting from the continuous parameters tuning, and a systematic approach to generate the FW BIC is still lacking. To address this, a method of actively forming FW BIC by matching the damping rate and resonance frequency of the coupling mode is proposed. As a proof-of-principle example, we propose a dielectric dimerized grating borophene heterostructure that generates a FW BIC near the commercially important communication wavelength. The coupling system comprises an electrically tunable borophene plasmon mode and a BIC supported by a dielectric dimer grating that can be attributed to the Brillouin zone folding. More interestingly, the BIC can be excited by the localized borophene plasmon (LBP) mode through near-field coupling as LBP mode can be considered as the dipole source. The interaction between them can further form the FW BIC, and support electromagnetically induced transparency (EIT)-like with maximum group index up to 2043, indicating its great potential for slow light applications. Our results provide a promising strategy and theoretical support for the generation of FW BIC in active plasmonic optical devices.

## I. INTRODUCTION

Optical trapping is essential in the interaction between light and matter, enabling the development of various optical devices and enhancing their performance. An optical cavity or resonator can confine light, but its efficiency is compromised when light couples to the propagating wave and leaks out if the frequency of the discrete mode falls within that of the continuum spectrum. The bound state in continuum (BIC) could be a viable solution to this issue. BICs are trapped for an infinite lifetime, even though they are embedded in the continuum spectrum of extended states. In 1929, John von Neumann and Wigner introduced the concept of BIC as a solution to the single-particle Schrödinger equation in quantum mechanics [1]. BIC has been applied in various systems, such as fluid dynamics, acoustics, electromagnetic waves, and water waves [2-11]. Recently, photonic nanostructures have become a suitable platform for investigating the properties of BICs, paving the way for potential applications in low-loss lasers, sensors, and filters [12-16]. Based on the intrinsic physical process, optical BICs with infinite lifetimes can be classified into several types, including symmetry-protected (SP) BICs [17-18], Fabry-Pérot (FP) BICs [19], and Friedrich-Wintgen (FW) BICs [20-22].

The incompatibility in symmetry between the incident wave of the continuum and the local mode typically results in SP BIC, while FP BIC is due to the destructive interference between two scattering channels [23]. According to Friedrich and Wintgen's theory, the interference of coupling between two resonances that radiate into the same channel can induce a vanished resonance called FW BIC [20]. This can be achieved by varying the structural parameters or excitation conditions. In recent years, researchers have focused on FW BIC because of its topology protection properties and potential application in on-chip tunable beam steering [24]. Extensive research has been carried out on the mechanism of optical BICs based on pure lossless dielectrics. The electromagnetic field is always loosely distributed in the structure, resulting in a large mode volume. The metallic structure supports a surface plasmon resonance mode that exhibits significant near-field enhancement and high mode confinement, despite absorption and scattering losses at visible and near-infrared wavelengths.

Researchers have recently studied the formation of BICs in systems with realistic intrinsic loss. A plasmonic FW BIC mode has been successfully formed in the visible wavelength range using an all-metallic grating by coupling the localized surface plasmon resonance and diffracted orders [25]. Azzam et al. proposed a hybrid plasmonic-photonic structure to realize FW BIC by coupling lossless photon modes to plasmon modes with ultra-mode confinement properties. This approach overcomes the limitations of the individual counterparts in hybrid plasmonic-photonic structures [26]. The FW BIC can be obtained by mode coupling when the resonance frequency and the radiative damping rate of the coupled modes are matched. Therefore, the development of effective methods to actively manipulate resonances with controllable resonant frequencies and radiative damping rates is a crucial and significant challenge. Very recently, an active metasurface was proposed to support plasmonic SP BIC by directly using graphene disk dimers as active atoms. The appearance of the BIC and its $Q$-factor can be tuned by applying a different external potential to each disk in the dimer [27]. This inspires the design of hybrid plasmon-photon systems from graphene-like materials with tunable conductivity, combining the advantages of all parties. However, plasmon resonances are not supported by graphene at commercially important communication bands. Borophene exhibits an ultrahigh electron density compared to other 2D materials, which enables support for tunable and deeply confined plasmon modes in the visible and near-infrared region [28-30].

Here, we investigate the active formation of FW BICs using a hybrid plasmonic-photonic structure. The structure comprises of a borophene plasmonic grating coupled to a dielectric dimerized grating with diverging radiative Q-factors. The dependence of BIC supported by the dielectric dimerized grating on the asymmetry parameter is calculated, and its formation mechanism could be attributed to the Brillouin zone folding. The quasi-BIC strongly coupled to the LBP mode leading to an avoided crossing behavior, gives rise to the emergence of FW BICs near an anti-cross point. More interestingly, the LBP mode could be considered as an electrically dipole source and then directly excite the BIC in the dielectric dimerized grating through the near-filed coupling. The interaction between them can further form the FW BIC, and support EIT-like with maximum group index up to 2043, indicating its great potential for slow light applications. Our results provide a promising strategy and theoretical support for the generation of FW BIC in active plasmonic optical devices.

## II. THEORY

To illustrate the physical mechanism of the FW BIC, the coupled mode theory (CMT) is employed to describe a system that involves two resonant modes with near-field and far-field coupling [31, 32]. The governing equation could be written by two mode amplitude $\mathbf{a}^\mathrm{T} = (a_1, a_2)$, effective Hamiltonian matrix $\mathbf{H}$, scattering matrix $\mathbf{C}$ and input (output) waves $s_+$ ($s_-$) with coupling coefficient $\mathbf{D} = (d_1, d_2)$ as

$$\frac{d\mathbf{a}}{dt} = -i\mathbf{H}\mathbf{a} + \mathbf{D}^\mathrm{T} s_+ = -i[\mathbf{\Omega} - i(\mathbf{\Gamma}_r + \mathbf{\Gamma}_n)]\mathbf{a} + \mathbf{D}^\mathrm{T} s_+ \tag{1}$$

$$s_- = \mathbf{C} s_+ + \mathbf{D}\mathbf{a} \tag{2}$$

The matrices are given as follows:

$$\mathbf{\Omega} = \begin{pmatrix} \omega_1 & g \\ g & \omega_2 \end{pmatrix}, \mathbf{\Gamma}_r = \begin{pmatrix} \gamma_{r1} & \sqrt{\gamma_{r1}\gamma_{r2}} \\ \sqrt{\gamma_{r1}\gamma_{r2}} & \gamma_{r2} \end{pmatrix}, \mathbf{\Gamma}_n = \begin{pmatrix} \gamma_{n1} & 0 \\ 0 & \gamma_{n2} \end{pmatrix}. \tag{3}$$

Here, $\omega_{1,2}$, $\gamma_{r1,2}$ and $\gamma_{n1,2}$ are the eigen-frequency, the radiative and dissipative damping rate of two mode, respectively, and $d_{1,2}$ denoting the coupling between the external radiation and two modes through the port. The off diagonal-terms $g$ and $\sqrt{\gamma_{r1}\gamma_{r2}}$ are the near-field and far-field coupling, respectively. It should be noted that the far-field coupling terms are deduced by the energy conservation and time-reversal symmetry. So the effective Hamiltonian matrix of the coupled system could be described as

$$\mathbf{H} = \begin{pmatrix} \omega_1 & g \\ g & \omega_2 \end{pmatrix} - i \begin{pmatrix} \gamma_{r1} + \gamma_{n1} & \sqrt{\gamma_{r1}\gamma_{r2}} \\ \sqrt{\gamma_{r1}\gamma_{r2}} & \gamma_{r2} + \gamma_{n2} \end{pmatrix}. \tag{4}$$

In order to correctly predict the positions of BIC, the appearance of BIC is analyzed as follows. By solving $|\mathbf{H} - \omega \mathbf{I}| = 0$, the eigenvalues for $H$ are as follows

$$\omega_\pm = \frac{\omega_1 + \omega_2}{2} - i\frac{\gamma_1 + \gamma_2}{2} \pm \frac{1}{2}\sqrt{[(\omega_1 - \omega_2) - i(\gamma_1 - \gamma_2)]^2 + 4(g - i\sqrt{\gamma_{r1}\gamma_{r2}})^2}, \tag{5}$$

where $\gamma_1 = \gamma_{r1} + \gamma_{n1}$, $\gamma_2 = \gamma_{r2} + \gamma_{n2}$, and $I$ being the identity matrix. The condition for the

FW BIC to be one of the solutions can be derived from Equation (5) by assuming that one of the solutions is purely real. On assuming $\gamma_{n1} = \gamma_{n2} = 0$ for simplicity, we obtain the Friedrich-Wintgen condition

$$g(\gamma_{r1} - \gamma_{r2}) = \sqrt{\gamma_{r1}\gamma_{r2}}(\omega_1 - \omega_2). \tag{6}$$

Hence, the eigenvalues could be written as

$$\omega_+ = \frac{\omega_1 + \omega_2}{2} + \frac{g(\gamma_{r1} + \gamma_{r2})}{2\sqrt{\gamma_{r1}\gamma_{r2}}} - i(\gamma_{r1} + \gamma_{r2}) \tag{7}$$

$$\omega_- = \frac{\omega_1 + \omega_2}{2} - \frac{g(\gamma_{r1} + \gamma_{r2})}{2\sqrt{\gamma_{r1}\gamma_{r2}}} \tag{8}$$

It is evident that the BIC is obtained at $\omega_1 = \omega_2$ when $g = 0$ or when $\gamma_{r1} = \gamma_{r2}$, indicating FW BICs near the frequency crossings of the uncoupled resonances.

## III. SIMULATION AND DISCUSSION

A. Exploring the BIC in dielectric dimerized grating and borophene plasmon mode

Fig. 1(a) shows a dimerized grating composed of two identical nanowires, which is composed of silicon with index $n = 3.48$. When no perturbation $\delta$ is introduced, the transverse distance $d$ between the nanowires in one unit cell is the same as the interspace between the unit cell's edges. The detailed geometry parameters are as follows, the period $P$ is 1000 nm, the height $h$ is 150 nm, the width $w$ is 300 nm, and the transverse distance $d$ is 200 nm without the perturbation $\delta$. The background index is set to 1 to simplify the simulation. The numerical simulations are performed by the finite-difference time-domain (FDTD) method with a 2D simulation region. The mesh is 2 nm in $y$ and $z$ direction to ensure accuracy. The $y$-polarized input waves are incident in the $z$-direction. The periodic boundary condition is used in the $y$-direction, and the perfectly matched layers (PMLs) boundary is used in the $z$-direction.

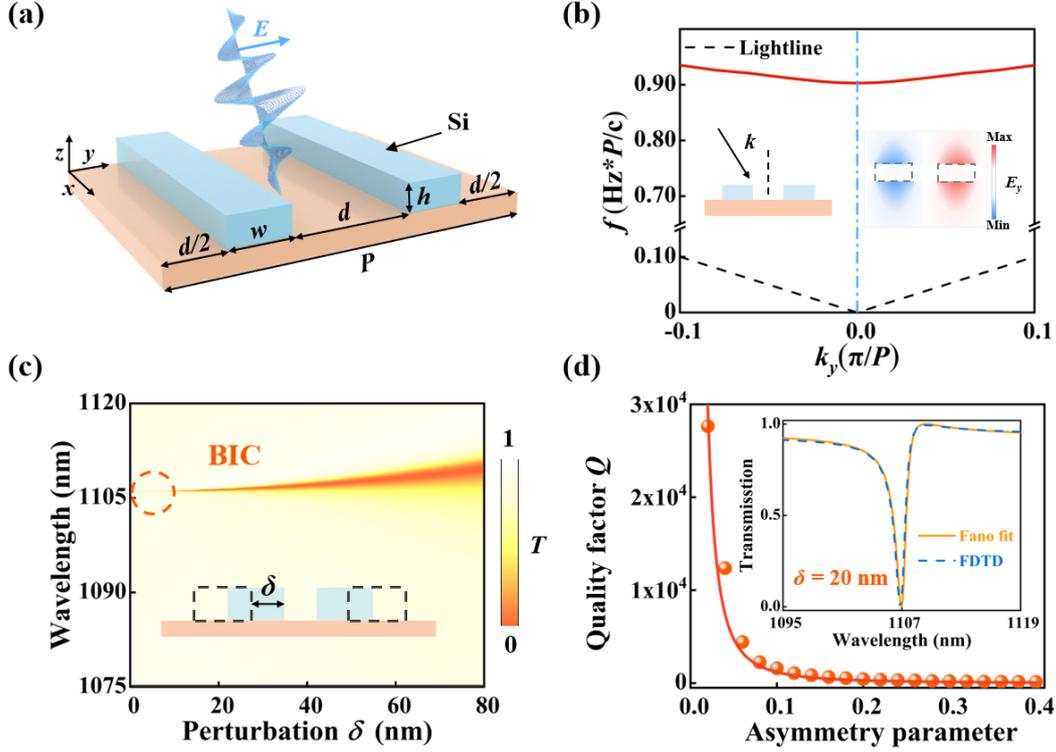

**FIG. 1.** (a) The schematic and (b) band structure of the dielectric dimerized grating. The inset is the field distribution $E_y$ of the BIC. The dashed line denotes the location of the nanowires. (c) The transmission spectra with varying perturbation $\delta$. The inset is the schematic of introducing perturbation. (d) The $Q$-factor as a function of asymmetry parameter. The inset is the transmission spectrum with $\delta$ = 20 nm.

To investigate the underlying physical mechanism of the BIC, the bandstructure is calculated in Fig. 1(b), where the BIC is excited through near-filed coupling by randomly placing dipole sources around the structure. The resonance wavelength of BIC is 1106 nm, and the near field distribution $E_y$ indicates that it concentrates around high-dielectric material rather than the air gap. When the perturbation $\delta$ is introduced, the BIC will transfer into quasi-BIC with finite quality factor $Q$ and non-zero linewidth, and its formation mechanism could be attributed to the Brillouin zone folding [6]. The transmission spectra with varying perturbation $\delta$ are presented in Fig. 1(c), where the linewidth broadens with the increase of $\delta$, corresponding to the more energy radiating into the continuum. The quasi-BIC exhibits a sharp Fano line shape in the transmission spectrum, as an example $\delta$ = 20 nm, shown in the inset of Fig.

1(d). The damping rate can be obtained by fitting the Fano formula [33]

$$T(\omega) = T_{bg} + T_0 \frac{[q + 2(\omega - \omega_0)/\gamma]^2}{1 + [2(\omega - \omega_0)/\gamma]^2}, \quad (9)$$

where $\omega_0$ is the resonance frequency, $\gamma$ is the damping rate, $q$ is the Fano asymmetry parameter, $T_{bg}$ is the contribution from continuum to resonant amplitude and $T_0$ is the offset. The fitting curve agrees well with the simulation, and the damping rate of quasi-BIC is 1.118 meV. Due to lossless material, the damping rate is totally contributed by the radiative term, i.e., $\gamma_r$ = 1.118 meV. In our case, the asymmetry parameter is defined as the $\delta/d$ to induce the BIC transferring the quasi-BIC. As shown in Fig. 1(d), the Q-factor of the quasi-BIC drops sharply with increased asymmetry parameter.

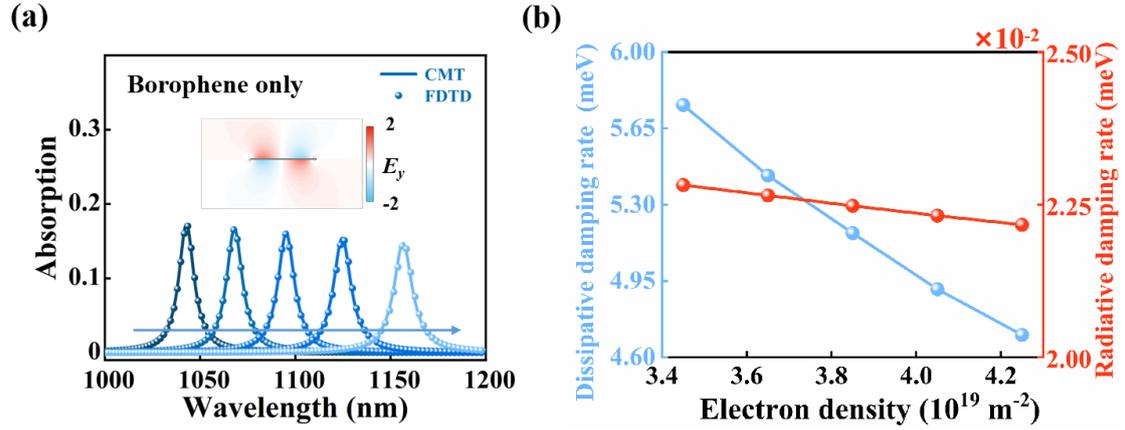

**Fig. 2.** (a) The absorption spectra of borophene plasmon mode with varying electron density $n$. The inset is the field distribution $E_y$ of the plasmon mode with $n = 4.25 \times 10^{19}$ m$^{-2}$. The line is the schematic of borophene. (b) The dependence of dissipative and radiative damping rate on electron density $n$.

The borophene plasmon mode with electrically tunable resonance wavelength offers a flexible way to manipulate the frequency detuning. The complex permittivity of the borophene is described as [34]

$$\varepsilon_{r,jj} = \varepsilon_d - \frac{e^2 n}{m_j \varepsilon_0 t(\omega^2 + \frac{1}{\tau^2})}, \varepsilon_{i,jj} = \frac{e^2 n/\tau}{m_j \varepsilon_0 t(\omega^2 + \frac{1}{\tau^2})}. \quad (10)$$

Here, $\varepsilon_d$ = 11 is the direct current permittivity of boron, $t$ = 0.3 nm is the thickness of the borophene, $\tau = 65$ fs is the electron relaxation time, $e$ is the electron charge, $m_j$ is the effective electron mass of optical axes, and $n$ is the electron density which could be electrically tuned.

As the absorption spectra demonstrated in Fig. 2 (a), when the electron density $n$ decreases from 4.250 to $3.250 \times 10^{19}$ m$^{-2}$, the resonance wavelength is red-shifted from 1040 to 1156 nm. The field distribution $E_y$ around borophene as the inset demonstrates that the LBP mode is an electric dipole mode. To investigate the radiative damping rate of the borophene plasmon mode, the CMT theory of single mode is used. The absorption of single mode could be described as [35]

$$A = \frac{P}{|S_+|^2} = \frac{2\gamma_r \gamma_n}{(\omega - \omega_0)^2 + (\gamma_r + \gamma_n)^2}. \tag{11}$$

The absorption spectrum (line) fitting by Eq. (11) aligns well with the simulation results (dots). The radiative and dissipative damping rates for a variation of the electron density $n$ are shown in Fig. 2 (b). The radiative terms vary slightly under the varied electron density $n$, while the dissipative terms change from 5.75 to 4.70 meV. When the wavelength of the borophene is adjusted to match that of quasi-BIC by manipulating the electron density, the radiative damping rate can be considered approximately constant. This could potentially lead to the achievement of FW BIC through the resonance frequency and radiative damping rate match.

B. The realization of Friedrich-Wintgen BIC

To verify the prediction from CMT, we propose the dielectric dimerized grating borophene heterostructures as shown in the inset of Fig. 3 (a). The width of borophene is $b = 160$ nm. In the case, $\delta = 2$ nm, the absorption spectra exhibit an anti-crossing behavior as shown in Fig. 3(a). Furthermore, the FW BIC has formed at 1106.7 nm with $n = 3.975 \times 10^{19}$ m$^{-2}$ (cycle mark) due to the interference of two modes. It should be noted that the radiative damping rate of two modes is not completely equal ($\gamma_{r1} \approx \gamma_{r2}$), so the FW BIC occurs near the zero-detuning frequency ($\omega_1 \approx \omega_2$). According to Eq. (8), the linewidth of the hybrid mode becomes narrower, i.e., the imaginary part becomes smaller in the upper band in the Fig. 3(a).

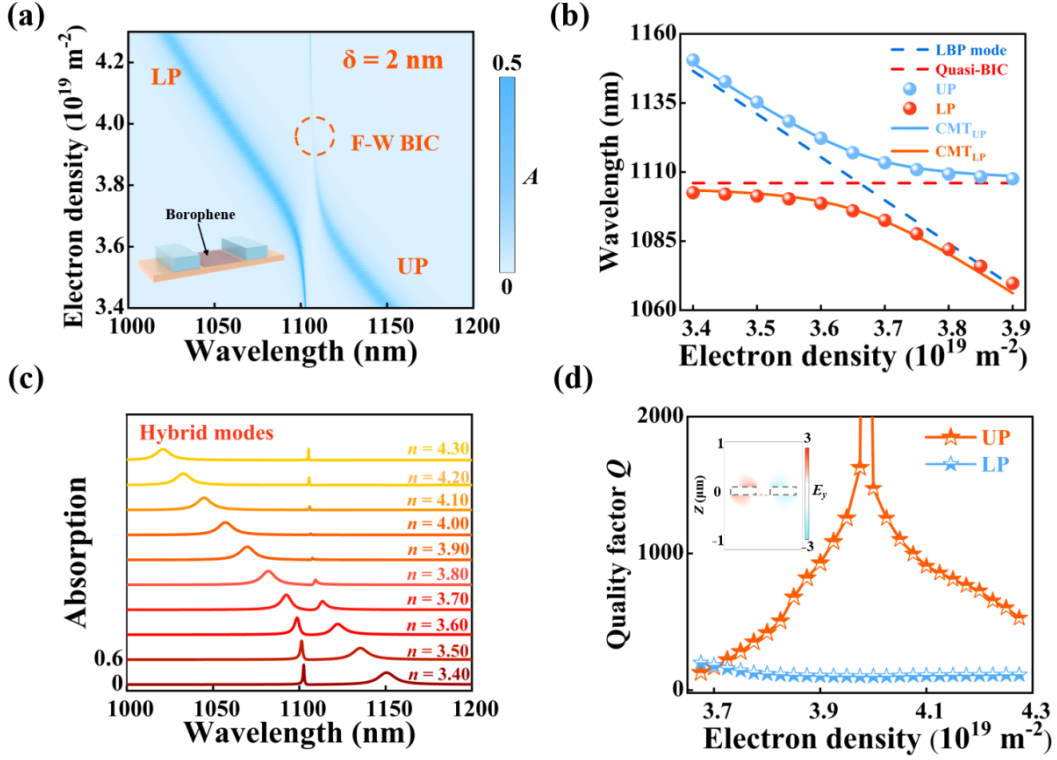

**Fig. 3.** (a) The absorption spectra with the perturbation $\delta = 2$ nm. The insert is the schematic of the coupled system. (b) The theoretical prediction of the wavelength of the proposed coupled system. (c) The simulated absorption with electron density $n$ varied from 2.40 to $4.30 \times 10^{19}$ m$^{-2}$ spectrum in (a). (d) The calculated quality factor of the lower and upper band as a function of electron density $n$. The inset is the field distribution $E_y$ of the coupled mode at the upper band with $n = 3.975 \times 10^{19}$ m$^{-2}$.

The resonance wavelength of the coupled mode predicted by Eq. (5) depicted in Fig. 3(b) is consistent with the simulation results. The damping rates of the quasi-BIC and LBP mode ($\gamma_1$ and $\gamma_2$) are calculated by fitting the spectrum, respectively. The resonance wavelength of LBP mode is approximately as [36]

$$\lambda_{LBP} = c\sqrt{\frac{4\pi m_j \varepsilon_0 \varepsilon_s b \xi}{ne^2}} \qquad (12)$$

Here, $\varepsilon_s$ is the equivalent permittivity of ambient, $b$ is the width of the sheet and $\xi$ is the fitting factor from the simulation results. The near-field coupling coefficient $g$ can be calculated via [37]

$$g = \frac{1}{2}\sqrt{R^2 + (\gamma_1 - \gamma_2)^2}, \tag{13}$$

where $R$ is the splitting energy attained by checking the distance between absorption peaks in simulation results. The evolutionary process of absorption spectra with varying $n$ is depicted in Fig. 3(c). Not only the linewidth of the hybrid mode is reduced, but the absorption amplitude around the BIC point. It can be explained by the ideal BIC which is decoupled to the continuum, resulting in it is not observed in the spectrum. The $Q$-factors is defined as

$$Q = \frac{\omega_0}{FWHM}, \tag{14}$$

where $\omega_0$ is the resonance frequency and FWHM is the full width half max of the resonance intensity spectrum, the $Q$-factors are calculated and depicted in Fig. 3(d). The $Q$-factor of the upper band can reach 1630 while the lower band is 105 under the electron density $n = 3.975 \times 10^{19}$ m$^{-2}$. The simulation results confirm that when the Friedrich-Wintgen condition satisfies, one mode becomes lossier, and the other is lossless. The field distribution $E_y$ at 1106.7 nm is shown in the inset of Fig. 3(d). The electric field distribution is characterized by the two coupled modes that the energy not only concentrates around the dielectric grating but the borophene. Under the case $\delta = 2$ nm, the BIC's occurring verifies the prediction of the CMT theory that the match of the radiative damping rate is the key point for the formation of FW BIC.

The structure with $\delta = 20$ nm while keeping the other geometry parameters unchanged, is also studied and the absorption spectra are shown in Fig. 4(a). When the resonance frequency of LBP mode crosses the quasi-BIC, the anti-cross behavior occurs, indicating the strong coupling between the quasi-BIC and LBP mode. However, due to the mismatch of radiative damping rate, the Friedrich-Wintgen condition is not satisfied. Under this condition, the imaginary part of the hybrid mode is approximately equal to each other when the detuning is zero. From the evolutionary process of absorption spectra with varying $n$ depicted in Fig 4(b), the linewidth of the hybrid modes are the same trend with the prediction of the CMT theory. Figure 4(c) shows that as the electron density $n$ increases, the $Q$-factors of the hybrid modes gradually cross each other. To provide insight into the formation of the FW BIC, Fig.

4(d) shows the radiative damping rates of the quasi-BIC and LBP mode for different perturbation $\delta$. The radiative damping rate of the quasi-BIC in dielectric dimerized grating is increased from 0.015 to 1.118 meV, while radiative damping rate of LBP mode ranges slightly. Hence, the radiative damping rates are mutually matched and then the FW BIC is achieved in the perturbation $\delta = 2$ nm rather than the $\delta = 20$ nm. With the guidance of this principle, the frequency of FW BIC can be extended to the other frequency band benefit from the scaling property of the dielectric metasurface.

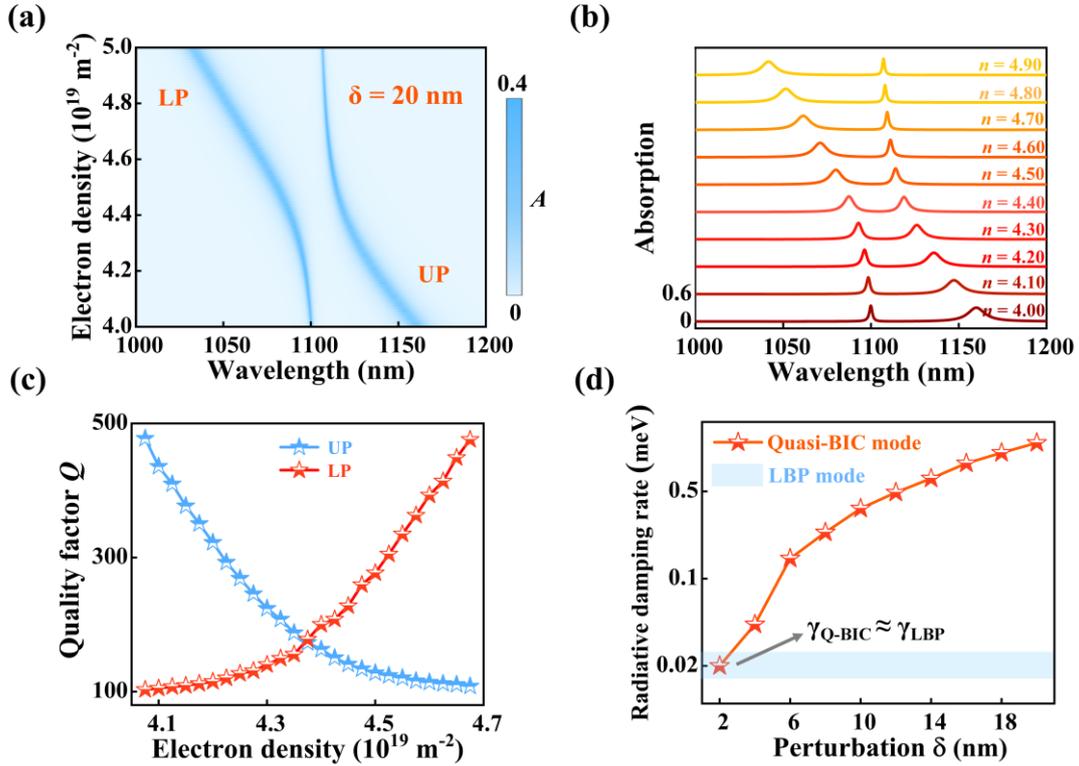

**Fig. 4.** (a) The absorption spectra with the perturbation $\delta = 20$ nm. (b) The extract absorption spectra with varying electron density $n$ varied in (a). (c) The calculated quality factors of the lower and upper band act as a function of electron density $n$. (d) The radiative damping rate of the quasi-BIC and LBP mode acts as a function of perturbation $\delta$, respectively.

C. The direct excitation of BIC by the mode coupling

Finally, we demonstrate that the BIC in the dielectric dimerized grating with $\delta = 0$ can be directly excited. In this case, the geometric parameters are the same as those in Fig. 1(a). As shown in Fig. 5(a), the LBP (bright) mode directly excited by the incident light can be considered as an electric dipole mode, it can excite the BIC (dark mode) by near-field coupling. The excitation of the BIC couples back to the LBP mode resulting in the energy splitting in the spectrum.

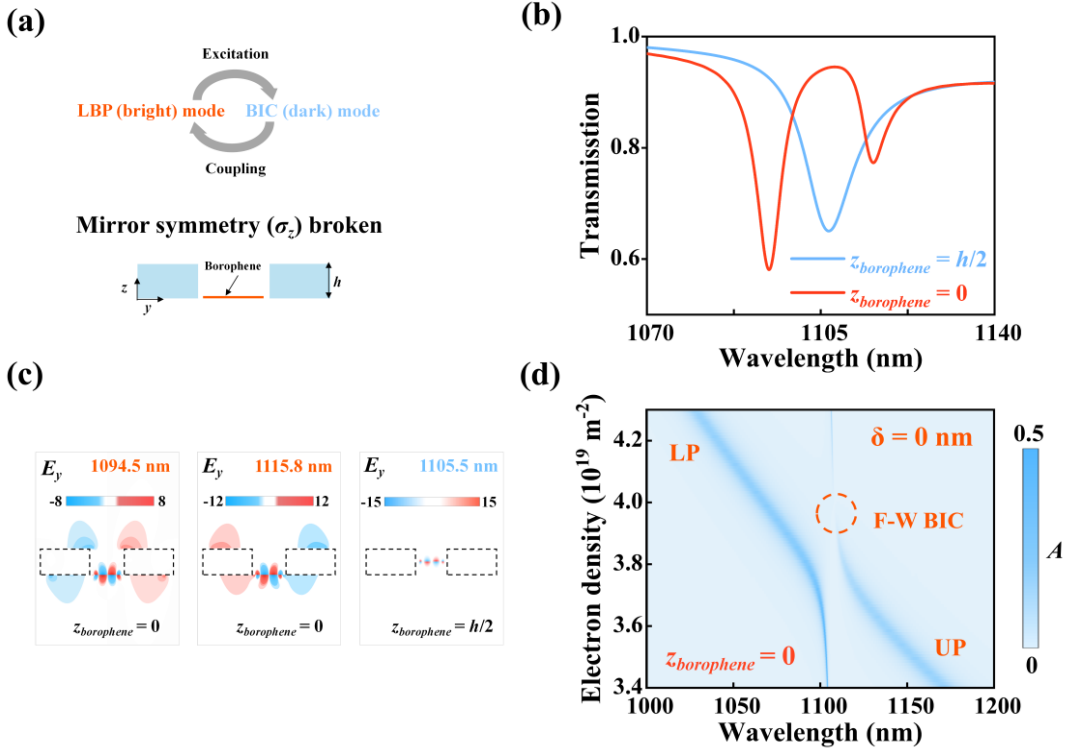

**Fig. 5.** (a) The illustration of excitation of BIC in the dielectric dimerized grating. (b) Transmission spectra with the borophene located at 0 and $h/2$, respectively. (c) The field distributions $E_y$ of the resonance dip in (b) at 1094.5, 1105.8 and 1105.5 nm, respectively. (d) The absorption spectra without the perturbation.

It should be noted that the excitation of the BIC depends not only on the near-filed coupling but the up-down symmetry broken of the system. As shown in the Fig. 5(b), when the longitude distance $z_{borophene}$ is zero, an obvious energy splitting occurs in the transmission spectrum due to the coupling as mentioned. The $E_y$ of two dips at 1094.5 and 1115.8 nm shown in Fig. 5(c) indicates that the BIC mode is indeed excited. Nevertheless, when the borophene locates at the $h/2$ in the $z$ direction, only one transmission dip at 1105.5 nm is observed in the spectrum. And the $E_y$ of the resonance is similar to the LBP mode. Moreover, as shown in Fig. 5(d), the FW BIC is also observed. It is intuitive that when the BIC is excited, it will radiate the energy to the continuum with a non-zero damping rate. So it's the damping rate also matched with the LBP mode.

To acquire deeper insights into the coupling between the BIC and the LBP mode, the transmission spectra with varying longitude distance of borophene from 50 to 100 nm is shown in Fig. 6(a). The resonance of LBP mode is approximatively unchanged due to the nearly-unchanged equivalent permittivity of ambient. The energy splitting obtained by checking the dips in the spectra is depicted in Fig. 6(b). The minimum and maximum of splitting energy are 10.185 and 1.379 meV corresponding to the distance 80 and 110 nm, respectively. The near-field coupling terms associated with the overlap of the electric fields of two modes can explain the change in splitting energy. The overlap of electric field density varies when the borophene is located at different longitudinal distances. The spectra show that the hybrid modes exhibit an EIT-like response. This effect can reduce the speed of propagation of electromagnetic waves and increase the interaction between light and matter. The group index $n_g$ is described by [38]

$$n_g = \frac{c}{l}\frac{d\varphi}{d\omega} \quad (15)$$

Here, the $\varphi$ is the phase of the transmission coefficient, and $l = 150$ nm is the height of the whole structure in $z$ direction. The phase and group index are shown in Fig. 6(c) with a distance 80 nm, and the maximum group index is up to 2043. The group index with varying distance is calculated and depicted in Fig. 6(d). The relationship between the group index and energy splitting is inversely proportional to the longitude distance. This is due to the fact that a smaller splitting energy leads to a more intense phase shift, resulting in a larger group index. Therefore, the proposed coupled system demonstrates significant potential in the field of slow light and could be useful in optical storage and other nonlinear applications.

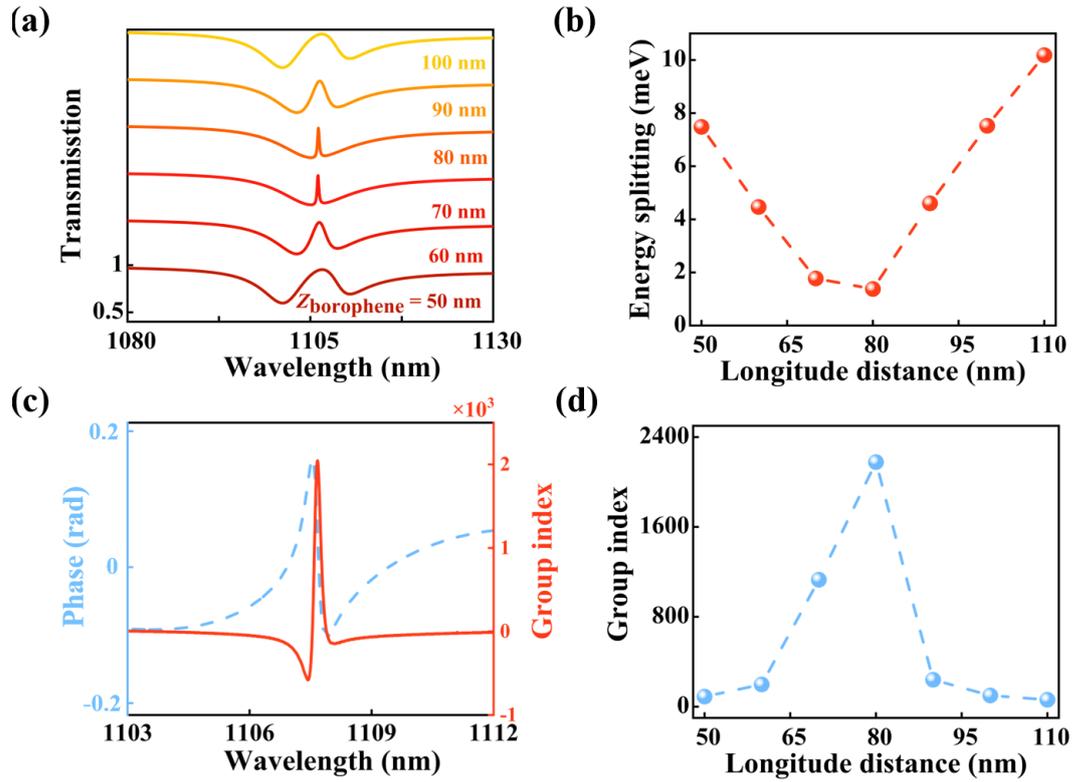

**Fig. 6.** (a) The transmission spectra with varying longitude distances of borophene sheet in $z$ direction. (b) The energy splitting as a function of longitude distance. (c) The phase and group index with the longitude distance = 80 nm. (d) The group index as a function of longitude distance.

In conclusion, we have demonstrated both numerically and theoretically the formation of FW BICs in a hybrid plasmonic-photonic structure consisting of a borophene plasmonic grating coupled to a dielectric dimerized grating with diverging radiative $Q$-factors. The dependence of BIC supported by the dielectric dimerized grating on the asymmetry parameter is calculated, and its formation mechanism could be attributed to the Brillouin zone folding. The quasi-BIC strongly coupled to the LBP mode leading to an avoided crossing behavior, gives rise to the emergence of FW BICs near an anti-crossing point. More interestingly, the LBP mode could be considered as an electrically dipole source and then directly excite the BIC in the dielectric dimerized grating through the near-field coupling. The interaction between them can further form the FW BIC, and support electromagnetically induced transparency with maximum group index up to 2043, indicating its great potential for slow light applications. Our results enhance the comprehension of FW BIC and offer theoretical backing for active plasmonic optical devices.

**Conflict of interest**

There are no conflicts to declare.

**Acknowledgements.**

This work is supported by the Scientific Research Foundation of Hunan Provincial Education Department (22B0105), Hunan Provincial Natural Science Foundation of China (2021JJ40523, 2020JJ5551), and the National Natural Science Foundation of China (62205278, 11947062, 62105276).